\documentclass[aps,showpacs,two column]{revtex4}
\usepackage{graphicx}
\usepackage[squaren]{SIunits}
\begin{document}

\title{Simulation of Current Experiments on the Tunneling Effect\\ of  Narrow $\varepsilon$-Near-Zero Channels}

\author{L. Zhao, Y. G. Ma, and C. K. Ong}
%\altaffiliation{Corresponding author} \email{ @nus.edu.sg }
\affiliation{Department of Physics, National University of
Singapore Singapore 117542
%Centre for Superconducting and Magnetic Materials, Department of
%Physics, National University of Singapore, 2 Science Drive 3,
%Singapore 117542, Singapore
}

\begin{abstract}
In this paper, we discussed the current experiments on the
tunneling effect of electromagnetic energy through narrow channels
of $\epsilon$-near-zero(ENZ) medium in the microwave range. Using
the finite element method, we carried out a full wave simulation
of the two kinds of experimental configurations at present. It was
shown that the ability of the electromagnetic waves penetrating
into the ENZ medium is very necessary. The present experimental
setups only using metamaterial-filled narrow channel in the
configuration of parallel plate waveguides are unlikely to realize
effective tunneling. Contrarily, the artificial plasma medium
emulated using hollow metallic waveguide can achieve nearly
perfect tunneling in a narrow channel without any transition
section around its cutoff frequency, which exceed the scope of
original ENZ tunneling theory and can be described by a simple
equivalent circuit model.

\end{abstract}
\vskip 15 pt 
\maketitle

\section{INTRODUCTION}
Since the rediscovery of the theory first proposed by Veselago in
1968\cite{veselago}, metamaterials have been an attracting subject
of growing worldwide interest. The metamaterials can possess many
exotic electromagnetic properties which are absent in natural
materials. So they are of particular importance in the engineering
of electromagnetic wave fields, which is leading to a variety of
new microwave or optical devices.

Most current research is still focused on the double negative
(DNG) metamaterials in which the permittivity and permeability can
both less than zero simultaneous, leading to the negative index of
refraction. Recently, there is a surge of investigation on
$\epsilon$-near-zero(ENZ) medium whose effective permittivity
$\epsilon$ is near zero and effective permeability $\mu$ remains
natural. In 2006, Silveirinha and Engheta\cite{engheta2006prl}
have theoretically predicted that the electromagnetic waves can be
squeezed and tunnel through very narrow subwavelength channels
filled with ENZ medium, independent of the specific geometry of
the channel. The incoming wave front can be replicated at the
output interface. The reflection at bends or junctions of
different devices can decrease greatly and even disappear using
this method. So, the ENZ medium will play an important role in the
applications of transport and interconnects of microwave and
optical wave energy.

Till now, several experiments have been reported to testify
Silveirinha and Engheta's ENZ theory and  the seemingly
'counterintuitive' tunneling effects are demonstrated at the
certain microwave frequencies\cite{liuprl, chengapl,
edwards2008prl,edwards2008arxiv}. The tunneling geometries in
these experiments are basically 'U-shaped'. At present, there are
two ways to experimentally realize the ideal ENZ tunneling
channels proposed by Silveirinha and Engheta.

In the work by Liu and Cheng et al\cite{liuprl, chengapl}, the
relatively complex artificial resonant inclusions implanted in the
parallel plate waveguides or microstrip circuits are adopted to
achieve the effective ENZ medium near the resonance frequency. The
two-dimensional(2D) scattering of the electromagnetic wave in the
original theory are simulated by the incident polarized nearly
transverse electromagnetic(TEM) waves.

In the later experiments, the other way was adopted  by Edwards et
al \cite{edwards2008prl,edwards2008arxiv}.  The artificial plasma
medium emulated by the TE$_{10}$ guided mode propagating in hollow
rectangular metallic waveguides, mimics the equivalent response of
ENZ materials near the cut-off frequency of guide wave modes.

In all above experiments the transition section in which the
incident mode ``penetrates" into the narrow channel is less
considered. Intuitively, the sharp discontinuities will cause
strong reflection of electromagnetic wave and excite high order
evanescent modes, which not considered in the discussion in
current work. To investigate this problem, we performed the full
wave simulations of these experiments respectively in the
following sections. From these results, we emphasis the importance
role of the transition section in the tunneling effect of the 2D
ENZ channel. The experiments results must be cautiously examined
to exclude other possibilities before being affirmed to the ENZ
tunneling phenomena predicted by the theory. While for the case of
hollow metallic waveguides, The effective tunneling can be
realized and the transition section is unnecessary. The
corresponding pictures of underlying physics are discussed.

\section{ENZ channel between parallel plate waveguides}

Over the past years, the development of metamaterials has greatly
 expanded the realm of available permittivity $\epsilon$ and
permeability $\mu$ \cite{smith2000}. The artificially
microstructured composites, may have become the most straight way
to realize the effective medium with certain electromagnetic
properties, including ENZ. In the work by Liu and Cheng
\cite{liuprl, chengapl}, the planar structure of complementary
split ring resonator (CSRR) are adopted, which have an electric
resonance in the certain range of frequency\cite{falcone}. The
resonance permittivity values of effective medium can be achieved.

\begin{figure}
\includegraphics[width=0.50\textwidth]{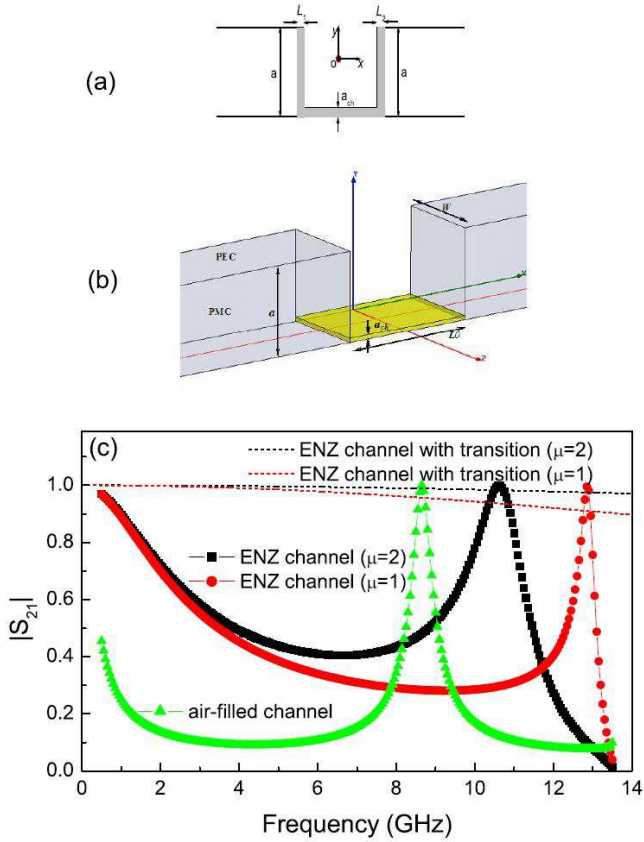}
\caption{ (a) A U-shaped geometry of the microwave tunneling
problem in a 2D parallel plate waveguide theoretically discussed
by Silveirinha and Engheta in Ref\cite{engheta2007prb}. (b) Setup
of our simulation, PMC boundaries conditions are on the x=0 and w
planes. Other outer faces are assigned as PEC boundaries.
$a=w=11$mm, and $a_{ch}$=1mm. (c) Simulation results of (b): the
transmission coefficients $|S_{21}|$ of ENZ channels with the same
geometry ($L_0=15mm$) and different permeabilities ($\mu/\mu_0=1,
2$). The one for a air-filled channel is also plotted. As a
comparison,  $|S_{21}|$ for the ENZ($\mu/\mu_0=1,2$) tunneling
model of the same size with side ENZ transition section
($L_1=L_2=a_{ch}$) is  plotted as dashed lines. \\}
\end{figure}

A general U-type geometry for the 2D tunneling problem of a ENZ
channel(shown in Figure 1(a)) has been discussed in detail as a
prototype in detail in Silveirinha and Engheta's
theory\cite{engheta2007prb}. Although contributing to the total
cross section ($A_p=(L_1a_1+L_2a_2)+La_{ch}$) and inconveniencing
the approximate condition ($k_0\mu_rA_p \rightarrow 0$ in
Ref\cite{engheta2006prl})to realize effective tunneling,  the
lateral ENZ sides is retained as transition section, allowing the
electromagnetic wave to penetrate into the narrow channel.

But in current experimental setup\cite{liuprl, chengapl}, only
channel filled with metamaterials are considered, without any
transition section of effective ENZ medium. The lateral  metallic
step sides are between parallel plate waveguides. Although in the
paper by Liu et al, they are considered as a shunt admittance in a
simple equivalent model\cite{collin}. But in their approximation,
it was oversimplified and can only be valid in a very limited
frequency range. The inhomogeneities at the metallic step in
waveguides are much more complicated. Furthermore, the incident
TEM-mode waves in parallel plate waveguide can excite more high
order evanesce modes, at each sharp metallic step, leading to the
interaction between these discontinuities, which can't be reckoned
in such a simple model. As we know, whether the tunneling effect
can be realized in this kind of configuration has not been
discussed theoretically yet.

Now we first consider the simplified ideal model, as shown in Fig.
1(b). Here the two planes, $x=0$ and $w$, are assigned as perfect
magnetic conducting(PMC) boundaries. Other outer faces except two
waveports are assigned as perfect electric conducting(PEC)
boundaries. These constraint conditions insure the incident waves
are of transverse electromagnetic(TEM) mode with certain
polarization as marked in (b). The field invariance along the z
direction can be fulfilled, so it degenerates to a 2D problem
independent of the width, $w$. If the effective medium theory
holds well for the volume with the CSRR or other composite
structures, this will be a ideal approximation for the experiments
by Liu and Cheng et al \cite{liuprl,chengapl}.

In our full wave simulation, the fixed height of the ENZ channel
$a_{ch}$(=0.5mm) is much less than the $y$ spacing of the input
and output parallel plate waveguides, $a=11$mm. Simulations of
this parallel waveguides geometry were performed with Ansoft's
High Frequency Structure Simulator (HFSS) based on finite element
method\cite{hfss}. Two set of constitutive parameters ($\mu_r$=1,
2) for lossless ENZ materials and are used.

For a fixed channel with $L0=15$mm, the simulated results of the
transmission coefficients $|S_{21}|$ are shown in Fig. 1(c)(the
lines with symbols). the air-filled channel without lateral are
simulated, and corresponding $|S_{21}|$ also plotted for
comparison.

The $|S_{21}|$ curves of the two ENZ channels without transition
section are far below unity in the most range of microwave
frequency, though they are above the level of air-filled channel.
For comparison, we simulated the ENZ tunneling model shown in Fig.
1(a) with side transition ENZ sections ($L_1=L_2=a_{ch}$). The
$|S_{21}|$ curves(dashed lines) suggest nearly perfect tunneling
($|S_{21}| \rightarrow 1$) through the the narrow channel with the
same size, independent of the permeabilities.

In the Fig. 1(b), the tangential components of electric fields
must be zero at the vertical surfaces of both side steps, which is
strongly non-compatible with the propagating TEM mode in the two
parallel plate waveguides. So the discontinuities obstacles the
propagation of electromagnetic waves for air or ENZ-filled
channel. On the contrary, the normal components of electric fields
must be zero due to finite electric flux density inside the ENZ
surface, while the tangential ones are continuous across the
interface. Thereby the lateral ENZ side layer let the normally
incident TEM waves penetrating into the ENZ directly with no need
of changing its configuration as shown in Fig. 1(a).

Another prominent character of the $|S_{21}|$ curves in the two
ENZ channels without transition sections is  the $|S_{21}|$ became
very significant(near unity)  around certain frequencies. The
position of transmission peaks in ENZ channels depends strongly on
the constitutive parameters, 12.8GHz i.e. and10.6GHz  for the ENZ
medium with $\mu$=1 and 2 respectively. The transmission peaks
also exist in air-filled channels at different frequencies. and
These have studied in detail as the Fabry-P$\acute{\texttt{e}}$rot
oscillation\cite{hibbins}. Does the peaks of $|S_{21}|$ in the ENZ
channels have the same origin? We further simulated six cases with
the different channels lengths, $L_0=10 to 60$mm with the same
permeability($\mu$=2). And the results are shown in Figure 2(a).

According the theory of Silveirinha and Engheta, the anomalous
tunneling effect is independent of geometry if only the ENZ
channel is narrow enough. Here,the frequency of these transmission
peaks increases as we decrease the length of channel section,
which shows a strong dependence of geometry.(The resonance peaks
also alter as we change the width of ENZ channel, $a_{ch}$(results
are shown here). So the origin of the tunneling effect can be
excluded.

To survey the underlying propagation processes in ENZ channels at
the resonance frequencies, the  distribution of the calculated
electric field at 0 phase  are  depicted in Figure 2(b) for a
typical channel with $L0=15$ at the corresponding resonance
frequency (10.6GHz). The magnitude of the calculated electric
field along the central line of the channel(red line in Figure
1(b)) is also plotted in the Fig 2(c). The
Fabry-P$\acute{\texttt{e}}$rot-like character of the strong
``standing" waves is observed in the ENZ channel with the maximum
magnitude at the two open ends of the channel, very similar to
Fabry-P$\acute{\texttt{e}}$rot oscillation in the air-filled slots
or channel\cite{hibbins}.

Commonly, the $m$th-order Fabry-P$\acute{\texttt{e}}$rot resonance
occurs at the frequencies near f =mc/(2nL0)(here c is the light
velocity in vacuum, h is the length) as  the channel width is
small enough. The transmission peaks will 'blue shift' as the
length channel increases, which is consistent with our results. So
these transmission peaks could be attributed to the
Fabry-P$\acute{\texttt{e}}$rot-like  resonance.

There is a non-zero minimum of the magnitude of field in the
middle of the channel, suggesting a slight difference from rigid
standing wave picture of classical Fabry-P$\acute{\texttt{e}}$rot
resonance. It may seem more confusing because the refraction index
($n= \sqrt{\epsilon \mu}$) of a ENZ medium is near zero, and the
wavelength of electromagnetic waves inside ENZ medium is
'infinite'. So no obvious phase difference should exist in ENZ
medium. The further thorough investigation on these anomaly
resonance peak is being on the way.

\begin{figure}
\includegraphics[width=0.50\textwidth]{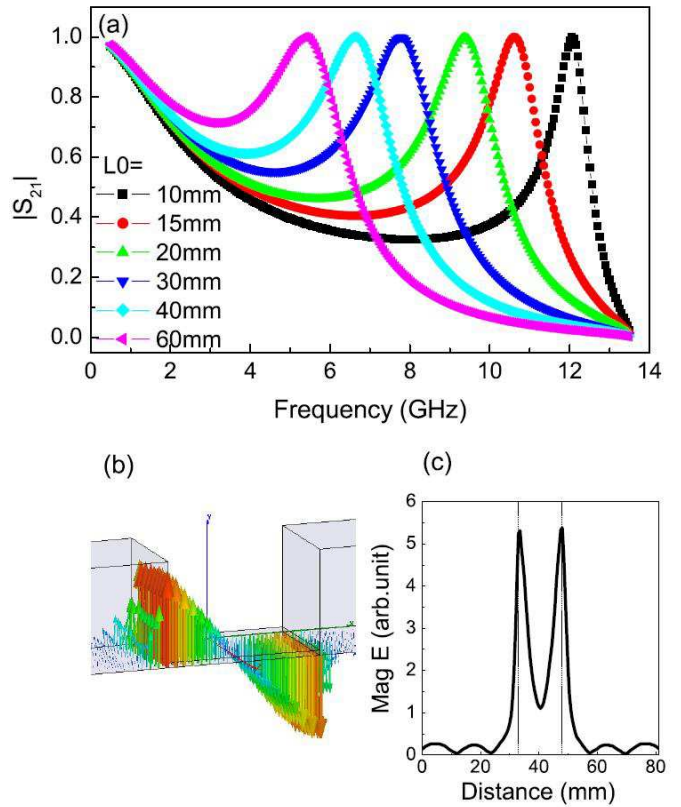}
\caption{(a) Simulated results of the transmission coefficients
$|S_{21}|$ of ENZ tunneling channels with different length $L_0$
at a fixed phase (b) Calculated  field distribution of electric
vector for the $L_0=15$mm ENZ channel at the corresponding
resonance frequency, 10.6GHz. (snapshot in time). The
corresponding magnitude of the electric field along the central
line of the channel are shown in (c).\\}
\end{figure}

These geometrical resonance characteristic of the transmission
peak is conceptually totally different from the tunneling effect
associated with ENZ channels which is independent of specific
geometry \cite{engheta2006prl}. So in practice researchers can
easily differentiate the resonance transmission peaks due to
either the ENZ tunneling effect or these
Fabry-P$\acute{\texttt{e}}$rot-like one by altering the
experimental geometry, such as the channel length $L_0$.

In these present experiments \cite{liuprl,chengapl}, the
inhomogeneities also exists at the metallic vertical steps and the
observed transmission $|S_{21}|$ at the passband is far below the
perfect tunneling level. The geometries of the tunnel channel were
fixed in their experiments. So the observed passband may not be
identified as the results of ENZ tunneling. The geometries of the
tunnel channel were fixed and never changed during their
experiments. Thereby their results may not be the strong evidence
of ENZ tunneling as they  declared.

\begin{figure}
\includegraphics[width=0.50\textwidth]{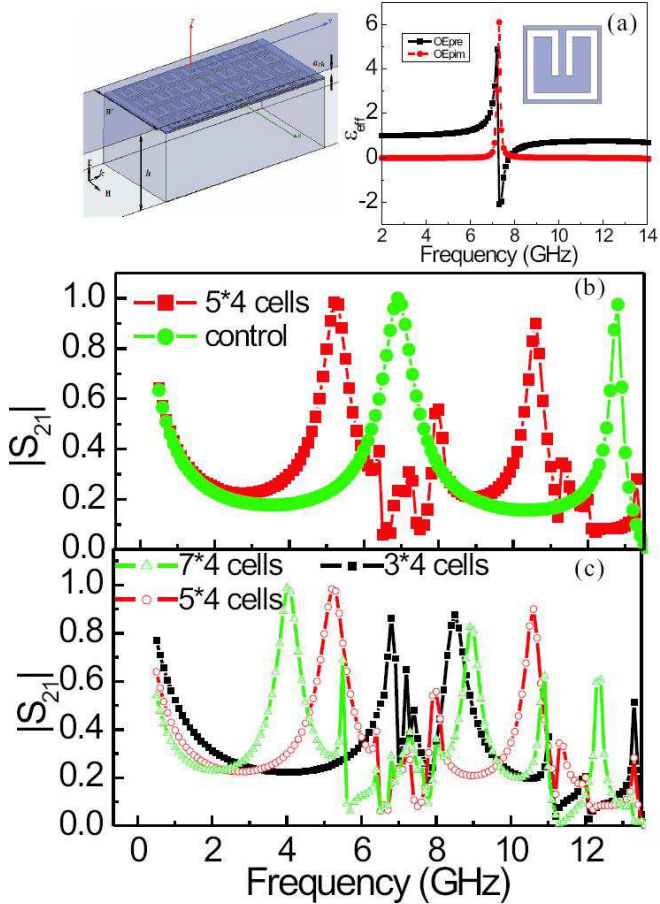}
\caption{ (a) Retrieved effective permittivity for a CSRR unit
cell(inset: the geometry of a CSRR cell). The simulation
configuration for the tunneling experiment in Ref\cite{liuprl} are
shown to the left of (a). The corresponding simulation results of
transmission coefficients $|S_{21}|$  as a function of frequency
are depicted in (b) and (c). Each tunnel channel consist of m
$\times$ n uniform CSRR cells (m cells in the propagation
direction and n cells in the transverse direction). The
$|S_{21}|$ for control sample is also plotted for comparison.
\\}
\end{figure}

A full wave $ab$ $initio$ simulation is necessary to give a more
clear picture. The configuration in our simulation (shown to the
left of Figure 3(a).) are nearly the same as the setup of the
tunneling experiment in Ref\cite{liuprl}, although with the less
size in transverse direction than adopted in experiments(about 40
cells) due to finite our limited computing resource. The details
of the CSRR cell structures(shown in the inset of Figure 3(a)) are
the same as in Ref\cite{chengapl, liuprl}, these CSRR patterns are
perfect metallic faces on the surface of FR4 substrate(0.2 mm
thick and $\epsilon /\epsilon_0 = 4.4$) in our setup. The
retrieval result of effective permittivity for one unit cell was
shown in the inset of Figure 3(a). The details of the retrieval
procedure are described in Ref\cite{liuprl, smith}.

In Figure 3(b), totally 5$\times$4 CSRR cells are used (5 in the
propagation direction and 4 in the transverse direction) to form
the effective ENZ medium channel, and the simulated transmission
$|S_{21}|$ vs frequency is depicted. The un-patterned copper-clad
FR4 board was used in the 'control' channel with the same
geometry, which was also simulated and corresponding $|S_{21}|$
plotted  for comparison. Apparently, the calculated $|S_{21}|$
curve for channel with CSRR patterns is similar to the control
one. But the resonance peak shifts about 2GHz to lower frequency
compared to the unpatterned control sample. There are weak fine
resonance structures between 6 and 8 GHz,  is consistent with the
electric resonance found in single CSRR cell retrieval procedure
(shown in Fig. 3(a)).

To find whether there exists  strong tunneling evidence in the
CSRR channel, we examined  CSRR channels with different
length($L_0$). Three cases, 3$\times$4, 5$\times$4, 7$\times$4,
are simulated, in which 3,5 and 7 cells are along the longitudinal
direction respectively. The corresponding results of  transmission
$|S_{21}|$ are depicted in Figure 3(c).

One can see that basically their behavior are similar and  the
peaks of the resonance transmission moves to low frequency as the
channel length increases, which agrees with the previous
Fabry-P$\acute{\texttt{e}}$rot-like oscillation discussed in
Figure 2(a). Apparently, the current geometry-sensible character
of the resonance transmission maximums in these CSRR patterned
channels didn't give any direct verification of the ENZ tunneling
theory. In current experiments using the configuration of TEM
parallel plate geometry, the transition section were neglected,
and the observed transmission peaks were more like to be  due to
Fabry-P$\acute{\texttt{e}}$rot-like oscillations. More work must
be done to improve the current experiments to realize genuine
verification of ENZ tunneling.

Additional, according to our results(shown in Fig. 3(c)), the fine
structures due to the resonance CSRR inclusion also shifts as vary
channel length.   The departure from the retrieving results from
one unit cell is very remarkable, suggesting the possible failure
of the effective medium in CSRR inclusions for finite size and its
inherent dispersion properties.

\section{Tunneling in hollow rectangular metallic waveguide operated at the
cut-off frequency)}

A close analogy has  been drawn between the propagating mode
within a waveguide and the plane wave propagation in equivalent
bulk medium with effective constitutive parameters\cite{rotman}.
Exploiting the inherent dispersion relation of hollow waveguides,
people have realized  plasma-like effective analogous to these of
metamaterials.

For the fundamental transverse electric (TE$_{10}$) mode in
rectangular metallic waveguide filled with non-magnetic medium,
phase constant is as follows:
$$\beta=\sqrt{(n\omega/c)^2-(\pi/w)^2 } =
\omega\sqrt{\mu_{0}\epsilon_{eff}}$$ where $\omega$ is the angular
frequency, $\mu_{0}$ the magnetic permeability of vacuum and n the
refraction index of medium filled in the waveguides. $w$ is the
fixed width of the waveguides. So by analogy, the effective
permittivity $\epsilon_{eff}$ can written as,
$$\epsilon_{eff} = \epsilon_0(n^2-c^2/(4f^2w^2))$$
c is the speed of light in air and f is the frequency. The cut-off
frequency $f_{0}=c\pi/w\sqrt{\epsilon}$ and the effective
permeability remains $\mu_{0}$, independent of the operating
frequency.

Thereby the TE$_{10}$ mode can be used to emulate the propagation
of a TEM wave in an classical lossless plasma medium. It has been
used in artificial plasma simulation at microwave frequencies in
Ref\cite{rotman}. In 2002, by inserting metamaterials with
effective negative permeability inside the rectangular waveguide
below the cutoff(i.e. $\epsilon_{eff}<0$), the equivalent  double
negative metamaterials have been realized.

In a series of recent work by Edwards et al\cite{edwards2008arxiv,
edwards2008prl}, many state-of-art demonstrations of nearly
perfect tunneling have been realized, using the dispersive
properties of rectangular metallic waveguides to mimic the
response of the ENZ medium. For the U-type geometry similar to the
experimental setup in Ref\cite{edwards2008prl}, our full wave
simulation using HFSS gives nearly the same results(dashed lines
shown in Fig. 4). Although the above simple equivalent analogy to
ENZ medium lacks of convincing theoretical evidence. It is
interesting that the configuration using rectangle waveguides can
meet well with the ideal ENZ theory so well. The nearly perfect
transmission around the cut-off frequency can be considered as a
nice circumstantial evidence of the equivalence of ENZ medium and
the waveguide.

Let's reconsider the origin derivation of their theory
\cite{engheta2006prl, engheta2007prb} on the tunneling through
narrow ENZ channel, the basic starting point is the z component of
H ($H_z$) holds as constant in the ENZ channel. So in another
word, if the $H_z$=const condition can be fulfilled in the narrow
channel in spite of specific filling, the tunneling effect can
sill occur without the precondition of ENZ medium.

In the TE$_{10}$ mode in lossless rectangle metallic waveguides,
the $H_z$ of TE$_{10}$ mode is written as\cite{duzer},
$$H_z=jA_{10}\frac{\beta w}{\pi}\sin(\frac{\pi}{w} x)e^{-\beta z}$$
where $A_{10}$ is constant coefficient and $\beta$ is the
propagation constant. Near the cut-off frequency, $\beta
\rightarrow 0$, and $\nabla H_z \rightarrow 0$, So the $ H_z=cons$
condition are satisfied automatically at its cut-off frequency. So
all the effects predicted by the tunneling theory
\cite{engheta2006prl, engheta2007prb}can be realized. And the
current present experiment can be thought as the more generic
demonstration of the original tunneling theory of Silveirinha and
Engheta.

From the discussion in Section II, without the transition section
(L1 and L2), an ENZ channel in  a parallel-plate geometry can't
lead to effective tunneling and the
Fabry-P$\acute{\texttt{e}}$rot-like resonance peaks appear at
certain frequencies instead . Is it the case for the hollow
waveguide configuration? To investigate this problem, we perform
the corresponding simulation with similar configuration without
lateral transition sections as shown in Fig. 1(b), in which two
PEC plates parallel to the x-y plane at  replaced the PMC ones as
boundaries at $x=0$ and $w$(we take $w= 20$ mm here). The narrow
channel are filled with air between two metallic rectangle
waveguides filled with teflon($\epsilon \sim 2\epsilon_0$),
supporting the propagation of TE$_{10}$ mode above its
corresponding cut-off frequency $f_0=c/2w\sqrt{\epsilon}\approx
5.3$GHz.

\begin{figure}
\includegraphics[width=0.50\textwidth]{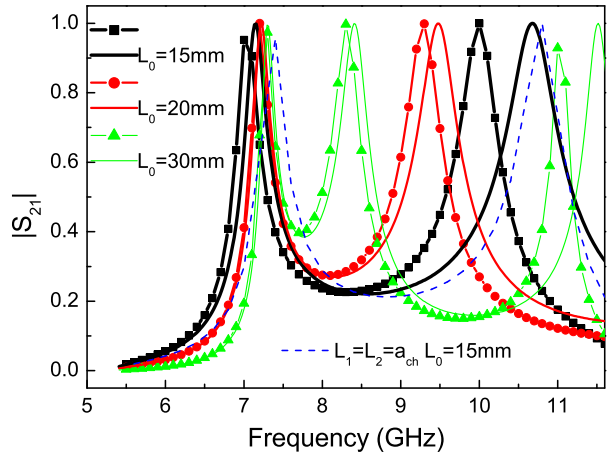}
\caption{ Results of transmission coefficients $|S_{21}|$ of
tunneling channels made of hollow metallic waveguides.   The model
is similar to Figure 1(b) with $a=10$mm, $w=20$mm, and
$a_{ch}=0.5$mm. All outer faces are PEC boundaries. Solid lines
with symbols are simulated $|S_{21}|$ for different channel
Lengths($L_0$=15, 20, 30mm) using HFSS. Similar solid lines
without symbols are corresponding results calculated from a
simplified equivalent circuit model(shown in Figure 6). The dashed
line is simulated $|S_{21}|$ of the channel modelled with lateral
transition sections($L_1=L_2=a_{ch}$).
\\}
\end{figure}

Our simulation results are depicted in Fig. 4. Near the cut-off
frequency of the air filled waveguide channel($f_0=c/2w=7.5$GHz),
it is surprising that the   $|S_{21}|$ curves for different
channel length($L_0=15-50$ mm) shows the peaks of the resonance
transmission around the cut-off frequency. The maximum of
corresponding peaks approaches unity, i.e. the nearly perfect
tunneling are achieved, independent of the geometry of the
channel. No remarkable difference found with the comparison case
with air-filled lateral transition section.

Although there are other transmission peak at higher frequencies,
their resonance frequency are strongly geometrically dependent and
Further analysis of corresponding field distribution(not shown in
this paper) indicate they can attribute to
Fabry-P$\acute{\texttt{e}}$rot resonances.

To confirm the tunneling nature of transmission peak near cut-off
frequency, the 3D field distribution of the electric field vectors
in a typical channel($L_0=15$mm and $a_{ch}=0.5$mm) at its
resonance frequency are plotted in Figure 5. The
TE$_{10}$-mode-distributed field are much stronger than those in
adjacent teflon filled waveguides, suggesting the huge squeezing
of electromagnetic energy by this air-filled narrow channel. In
the propagation direction, there is little change of the electric
field which the channel. Thereby the phase difference between two
end of the channel is near zero, as predicted in the ENZ tunneling
theory\cite{engheta2007prb}.

\begin{figure}
\includegraphics[width=0.40\textwidth]{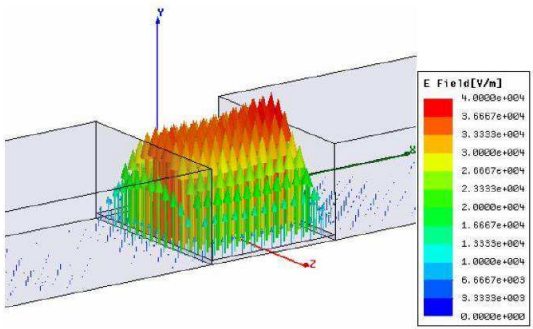}
\caption{ Calculated  field distribution of electric vector in the
rectangle waveguide tunneling channel with length $L_0$=15mm at
the resonance frequency near its cutoff.
\\}
\end{figure}

Only the narrow air-filled channel itself between two waveguides
can realize the effective tunneling of the incident TE$_{10}$
electromagnetic mode. It seems very surprising because of the very
different geometric and impedance mismatch between air-filled
channel and two teflon-filled waveguide regions. So it is
necessary to consider the E-plane steps discontinuities in
metallic waveguides, which is an classical problem which can be
simplified and analyzed using equivalent circuits model in
Ref\cite{wghandbook,rozzi}.  To understand our current results, we
modelled the whole system to be a simplified equivalent circuit
(shown in Figure 6).

The fundamental TE$_{10}$ mode is propagating in both these
waveguides with phase constants as follows,
$$\beta_i=\sqrt{\omega^2\mu_0\epsilon_i-(\pi/w)^2}$$

\begin{figure}
\includegraphics[width=0.40\textwidth]{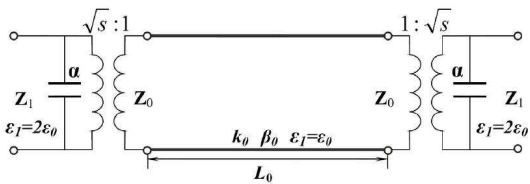}
\caption{Simplified circuit model for the metallic rectangle
waveguide channel.
\\}
\end{figure}

where the wave impedance is $ Z_{i}={\omega \mu_0}/{\beta_i}$
respectively for each section (i=0 for air-filled channel and 1
for teflon-filled waveguides). And an ideal
transformer\cite{rozzi} with the ratio $1:\sqrt{s}$ ($s=a_{ch}/a$)
was used in the model to reflect the change of impedance. The
numerical calculation have been described \cite{rozzi} in detail.
The calculated $S_{21}$ for different $L_0$ have been shown in
Fig. 4. The results  agree very well with those from the full wave
simulation using HFSS. The difference becomes slightly apparent
only at high frequency.

Around the cut-off frequency, the phase velocity became infinite
as $\beta \rightarrow 0$ and The change of E and H is along
longitudinal direction. So $L_0$ is negligible. On the other hand,
when $s \rightarrow 0$, the change of circuit impedance will be
dominated by the two transformers (with the ratio $1:\sqrt{s}$ and
$1:\sqrt{s}$) and relatively the effect of the shunt admittance
can be neglected. (In our calculation, the moderate change on the
value of the  admittance load, $\alpha$, have very weak effect on
the position of resonance transmissions.) The super-coupling
channel connected two waveguides with the same impedance can be
obtained. The excellent agreements suggest that such a simple
circuit model may grasp the nature of tunneling physics in
metallic rectangular waveguides. Further experimental
demonstrations of the tunneling in waveguides are expected in our
future work.

\section{CONCLUSION}

In summary, we systematically  investigated the recent experiments
about the demonstration of the tunneling effect through the narrow
ENZ channel. It is shown that the ability of the electromagnetic
waves penetrating into the ENZ medium is very necessary in the
parallel-plate setup, and the transition section is indispensable
to demonstrate effective tunneling effect. However for the
experimental configuration made of hollow metallic waveguides, the
nearly perfect tunneling can be achieved without transition
section near its cutoff frequency, which is a more generic case
and  exceed the scope of the original ENZ tunneling theory.

\clearpage
\end{document}